\documentclass[twocolumn,showpacs,preprintnumbers,amsmath,amssymb,showkeys]{revtex4}
\usepackage[T1]{fontenc}
\usepackage[utf8]{inputenc}
\usepackage{graphicx}
\usepackage{dcolumn}
\usepackage{bm}

\DeclareMathOperator{\Tr}{Tr}

\begin{document}

\title{Supercurrent dephasing by electron-electron interactions}
\author{Andrew G. Semenov$^{1,3}$
and Andrei D. Zaikin$^{2,1}$
}
\affiliation{$^1$I.E.Tamm Department of Theoretical Physics, P.N.Lebedev
Physics Institute, 119991 Moscow, Russia\\
$^2$Institute of Nanotechnology, Karlsruhe Institute of Technology (KIT), 76021 Karlsruhe, Germany\\
$^3$National Research University Higher School of Economics, 101000 Moscow, Russia
}

\begin{abstract}
We demonstrate that in sufficiently long diffusive superconducting-normal-superconducting ($SNS$) junctions dc Josephson current is
exponentially suppressed by electron-electron interactions down to zero temperature.
This suppression is caused by the effect of {\it Cooper pair dephasing}
which occurs in the normal metal and defines a new fundamental length scale $L_\varphi$ in the problem. Provided the temperature length exceeds $L_\varphi$ this dephasing length can be conveniently extracted from equilibrium measurements of the Josephson current.
\end{abstract}


\maketitle

\section{Introduction}
It is well established that hybrid metallic structures can sustain a non-vanishing
supercurrent even if they contain a non-superconducting region located in-between two
superconducting reservoirs. The physical reason for that is transparent: Cooper pairs passing through this region can maintain their macroscopic quantum coherence and, hence, their ability to carry a non-dissipative current through the whole structure.  This is the celebrated  dc Josephson effect which was initially predicted
for superconducting tunnel junctions \cite{BJ,AB} and later investigated
in other types of superconducting weak links, such as, e.g., quantum point contacts \cite{KO,HKR,Been} and superconductor-normal-metal-superconductor ($SNS$)
junctions \cite{ALO,SNScl,SNSc2,Likh,SNSd,KL,SNSd2,SNSd3,SNSc3}. In contrast to tunnel junctions containing insulating barriers with typical thicknesses of
few angstroms, in $SNS$ systems at sufficiently low temperatures appreciable supercurrent can flow even through  a normal layer as thick as few microns. The latter feature is generic in both limits of ballistic \cite{SNScl,SNSc2,SNSc3} and diffusive \cite{ALO,Likh,SNSd,KL,SNSd2,SNSd3} metals irrespective of the quality of $NS$-interfaces ranging from poor \cite{ALO,KL,SNSc3}
to perfect \cite{SNScl,SNSc2,Likh,SNSd,SNSd2,SNSd3}.
Quantitative agreement between theory and experiment was demonstrated in diffusive $SNS$ junctions with ideal \cite{SNSd3} and non-ideal \cite{SNSd4} $NS$-interfaces.
For a comprehensive coverage of this and other issues related to dc Josephson effect in different types of superconducting weak links we refer the reader to the review papers \cite{lam,bel,SaMiZhe}.

It is important to stress that all the above results apply provided the effect of electron-electron interactions remains weak and can be neglected.
However, the situation may be different in sufficiently small superconducting junctions in which case Coulomb effects can play an important role
and need to be taken into account. In the case of tunnel barriers between superconductors Coulomb blockade of Cooper pair tunneling results in a large
number of qualitatively new features which have been studied in a great detail \cite{SZ}.
With increasing barrier transmission Coulomb effects remain qualitatively the same though
decrease in magnitude and eventually vanish in the limit of fully open barriers between
superconducing electrodes \cite{ZGalakt3}. One can also consider the effect of Coulomb interaction on the Josephson current in more
complicated superconducting weak links, such as, e.g., diffusive $SNS$ junctions with low transmission $NS$-interfaces. For instance, the
authors \cite{BFS} addressed this problem within the so-called capacitance model assuming that Coulomb interaction merely occurs across
tunnel barriers at inter-metallic interfaces. They demonstrated that Coulomb effects result in effective reduction of the
Josepson current through the system.

In this work we will argue that electron-electron interactions also provide an alternative mechanism of the supercurrent suppression not directly related to
Coulomb blockade. It is {\it quantum dephasing of Cooper pairs} which yields exponential reduction of the Josephson current in sufficiently long diffusive $SNS$ junctions even in the zero temperature limit.

Note that previously various aspects of the effect of electron-electron interactions on dissipative (Andreev) currents in $NS$ hybrid structures were studied by a number of authors
\cite{Z,HHK,ZGalakt,ZGalakt2,SZK}. In particular, interaction-induced quantum dephasing in such structures was addressed both phenomenologically \cite{HHK} and
microscopically \cite{SZK} demonstrating that at sufficiently low temperatures this effect may strongly modify the Andreev conductance of $NS$ systems provided
the size of the normal metal becomes comparable with (or bigger than) the fundamental scale of dephasing length $L_\varphi$ which is set by interactions and stays finite down to
zero temperature. Although the effect of quantum dephasing of Cooper pairs by electron-electron interactions \cite{SZK} is to a large extent similar to that
previously investigated for normal electrons \cite{GZ1,GZ3,GZ4,GZS,GZ5} within the framework of the so-called weak localization problem, there are also
important differences between these effects in $NS$ structures and normal metals. They are caused, e.g., by the different spin structure of the
propagators describing Cooper pairs and single electrons in such systems \cite{SZK} as well as by some other features. Hence, it is not possible to
directly adapt the results \cite{GZ1,GZ3,GZ4,GZS,GZ5} to superconducting hybrids in which case a separate analysis is required. This analysis will be developed below
for the Josephson current flowing across diffusive $SNS$ junctions.

The structure of the paper is as follows. In section 2 we describe our theoretical approach based on the real time (Keldysh) version of the nonlinear $\sigma$-model. In section 3 this approach is employed for the analysis of the dc Josephson current in diffusive $SNS$ structures in the presence of electron-electron interactions. The effect of interaction-induced Cooper pair dephasing on the supercurrent is addressed in section 4. The paper is concluded by a discussion of our key observations in section 5. Further technical details are relegated to Appendices A and B.

\section{The model and basic formalism}
\begin{figure}[t]
\includegraphics[width=0.95\columnwidth]{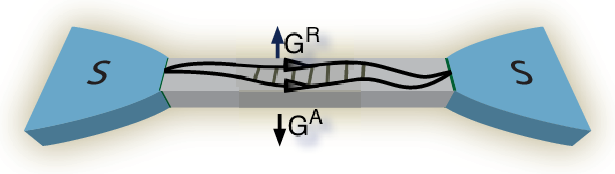}
\caption{Diffusive $SNS$ Josephson junction. The figure also illustrates the Cooperon and its spin structure relevant for the supercurrent flowing across the junction.}
\end{figure}

Let us consider an $SNS$ structure depicted in Fig. 1 illustrating two bulk superconducting leads connected by a normal wire of length $L$ and cross-section $a^2$.
In what follows we will merely stick to the limit $L \gg a$ and assume that superconducting electrodes are sufficiently large, i.e.
they are not influenced by the central (normal) part of our system. The normal wire is characterized by the density of states per spin $\nu$ and
diffusion coefficient $D=v_F \ell /3$, where $v_F$ is the Fermi velocity and $\ell$ is the electron elastic mean free path. The left and right superconductors
are connected to the normal wire via tunnel barriers with resistances $R_{l}$ and $R_{r}$ which are assumed to strongly exceed the wire normal resistance, i.e.
$R_l,R_r\gg R_N=L/(\sigma_N a^2)$, where $\sigma_N =2e^2\nu D$ is the Drude conductivity, and $e$ is the electron charge.

Provided the superconducting phase twist $\theta$ is applied to this $SNS$ structure
it develops a supercurrent $I$ which is a $2\pi$-periodic function of $\theta$.  The task at hand is to evaluate this supercurrent
in the presence of electron-electron interactions.

In order to accomplish this goal we employ a real-time version of the non-linear $\sigma$-model approach which provides
an effective low-energy description of disordered metals where the relevant degrees of freedom are diffusive collective modes, the so-called
diffusons and Cooperons. The information about these modes is contained in the $4\times 4$ matrix (in both Keldysh and Nambu spaces)
dynamical variable $\check Q$ which  depends on the spatial coordinate and two times, i.e. $\check Q=\check Q({\bm r},t,t')$.

The effective action of our system consists of three parts $S=S_w+S_I+S_\Phi$. The first two terms
account respectively for diffusive motion of electrons inside the wire,
\begin{equation}
S_{w}[\check Q,{\bm A},\Phi]=\frac{i\pi\nu}{4}\Tr[D(\check \partial\check Q)^2-4\check\Xi\partial_t\check Q+4i\check\Phi\check Q],
\label{sw}
\end{equation}
and electron tunneling between the wire and the leads \cite{KL},
\begin{equation}
S_I[\check Q]=-\frac{i\pi}{4e^2R_ra^2}\Tr_r[\check Q_{SC}^r\check Q]-\frac{i\pi}{4e^2R_la^2}\Tr_l[\check Q_{SC}^l\check Q],
\label{si}
\end{equation}
while the third term $S_\Phi[\Phi]$ is responsible for electron-electron interactions.
Here the matrix $\check Q^{l(r)}_{SC}$ accounts for the left (right) bulk superconducting electrode and, hence, is independent of the matrix $\check Q$.

The covariant derivative in Eq. (\ref{sw}) is defined as
\begin{equation}
\check\partial \check Q=\partial_{{\bf r}}\check Q-i[\check\Xi\check {\bf
A},\check Q], \quad \check\Xi=\left(\begin{array}{cc} \hat \sigma_z &  0 \\
0 & \hat\sigma_z \end{array}\right),
\end{equation}
where $[x,y]$ is the commutator and the set of Pauli matrices here and below is denoted by $\hat\sigma_{x}$, $\hat\sigma_{y}$ and $\hat\sigma_{z}$. The dynamical variable $\check Q$ satisfies the standard normalization condition
\begin{equation}
\check Q^2=\check 1\delta(t-t').
\end{equation}
All multiplications in the above expressions are meant as convolution of matrices (implying the integration over intermediate times) and
$\Tr$ indicates the trace over the matrix indices accompanied by the integration over both time and coordinate variables.
Note that below we will also employ a special multiplication notation defined as
\begin{equation}
(A\circ B)(t_1,t_2)=\int dt A(t_1,t)B(t,t_2).
\end{equation}

The effective action $S$ of our system depends on the scalar and vector potential fields
$\Phi({\bf r},t)$ and ${\bf A}({\bf r},t)$ which account for the effect of electron-electron interactions. These potentials are defined on the forward
($F$) and backward ($B$) branches of the Keldysh contour. For our purposes it is convenient to introduce the variables
$\Phi^{\pm}=\frac{1}{\sqrt{2}}(\Phi^F\pm\Phi^B)$ and  ${\bf
  A}^{\pm}=\frac{1}{\sqrt{2}}({\bf A}^F\pm{\bf A}^B)$ and to define the matrices
\begin{equation}
\check\Phi =\left(\begin{array}{cc} \Phi^+\hat1 & \Phi^-\hat1  \\
 \Phi^-\hat1 & \Phi^+ \hat1
 \end{array}\right), \quad\check{\bm A}=\left(\begin{array}{cc} {\bm A}^+\hat1 & {\bm A}^-\hat1  \\
 {\bm A}^-\hat1 & {\bm A}^+ \hat1
 \end{array}\right).
\end{equation}

Similarly to our earlier works \cite{ZGalakt,SZK} we will employ the so-called $\mathcal K$-gauge trick \cite{KA,KA1} and perform the gauge transformation
\begin{equation}
\check Q({\bm r},t,t')\to e^{i\check\Xi \check{\mathcal K}({\bm r},t)}\check Q({\bm r},t,t') e^{-i\check\Xi \check{\mathcal K}({\bm r},t')}
\end{equation}
in order to eliminate the linear terms in both the electromagnetic potentials and deviations from the normal metal saddle point
\begin{gather}
\check Q_N=\check {\mathcal U}\circ \left(\begin{array}{cc}
\hat\sigma_z&0\\0&-\hat\sigma_z
\end{array}\right)\check {\mathcal U},
\label{nsp}
\\
\check{\mathcal U}(t,t')=\left(\begin{array}{cc}\delta(t-t'-0)\hat 1 &
F(t,t')\hat 1 \\ 0 &
-\delta(t-t'+0)\hat1\end{array}\right),
\end{gather}
where
\begin{equation}
F(t,t')=\int \frac{d\varepsilon}{2\pi}e^{-i\varepsilon t}\tanh\left(\frac{\varepsilon}{2T}\right)=-\frac{iT}{\sinh(\pi T(t-t'))} .
\end{equation}
The latter goal is achieved if one chooses the $\mathcal K$-field to obey the following equations
\begin{eqnarray}
 \Phi_{\mathcal K}^+({\bm r},t)&=&D\partial_{\bm r}{\bm A}_{\mathcal K}^+({\bm r},t) \\
&&-2iDT\int dt'\coth(\pi T(t-t'))\partial_{\bm r}{\bm A}_{\mathcal K}^-({\bm r},t'),\nonumber\\
 \Phi_{\mathcal K}^-({\bm r},t)&=&-D\partial_{\bm r}{\bm A}_{\mathcal K}^-({\bm r},t)
\end{eqnarray}
with $\Phi_{\mathcal K}({\bm r},t)=\Phi({\bm r},t)-\partial_t\mathcal K({\bm r},t)$ and ${\bm A}_{\mathcal K}({\bm r},t)={\bm A}({\bm
r},t)-\partial_{\bm r}\mathcal K({\bm r},t)$. As a result of this transformation the total action retains its initial form provided one substitutes
$\Phi\to\Phi_{\mathcal K}$ and ${\bm A}\to{\bm A}_{\mathcal K}$ as well as
\begin{equation}
\check Q_{SC}^{l(r)}(t,t')\to e^{-i\check\Xi \check{\mathcal K}({\bm r},t)}\check Q_{SC}^{l(r)}(t,t')e^{i\check\Xi \check{\mathcal K}({\bm r},t')}
\end{equation}
with coordinate $\bm r$ chosen at the appropriate tunnel barrier at the $NS$-interface.

Finally, let us define the matrices $\check Q_{SC}^{l}$ and $\check Q_{SC}^{r}$ describing respectively the left and the right
superconducting electrodes. The first of these matrices reads
\begin{equation}
\check Q_{SC}^{l}=\check Q_{SC}\equiv\check{ \mathcal U}\circ\left(\begin{array}{cc}
\hat G^R & 0 \\
0 & \hat G^A
\end{array}
\right)\circ \check{ \mathcal U},
\end{equation}
where
\begin{multline}
 \hat G^R(t,t')=\int\frac{ d\varepsilon }{2\pi}\frac{e^{-i\varepsilon(t-t')}}{\sqrt{(\varepsilon+i0)^2-\Delta^2}}\left(
 \begin{array}{cc}
  \varepsilon & \Delta \\
  -\Delta & -\varepsilon
 \end{array}
 \right)
 \\\\\equiv
 \left(\begin{array}{cc}
 g_R(t,t') &f_R(t,t') \\
 -f_R(t,t') & -g_R(t,t')
 \end{array}\right)
=\delta(t-t'-0)\hat\sigma_z\\-
 \theta(t-t')\Delta\left(\begin{array}{cc}
 J_1(\Delta(t-t'))&iJ_0(\Delta(t-t'))\\
 -iJ_0(\Delta(t-t'))&-J_1(\Delta(t-t'))
 \end{array}
 \right),
  \end{multline}
 $J_k(x)$ is the $k-$th Bessel function and $\hat G^A=-[\hat G^R]^T$ with the transposition $[...]^T$ performed in both matrix indices and times.
For the second (right) electrode one has
\begin{equation}
\check Q_{SC}^r(t,t')=e^{i\check\Xi(\theta+\check\Upsilon \chi(t))/2}\check Q_{SC}(t,t')e^{-i\check\Xi(\theta+\check\Upsilon \chi(t'))/2},
\end{equation}
where
\begin{equation}
\check\Upsilon=\left(\begin{array}{cc}
0 &\hat 1\\
\hat 1 & 0
\end{array}
\right),
\end{equation}
$\theta$ is the superconducting phase difference between the two electrodes and $\chi(t)$ is the source field. Taking the variation over this source field
as
\begin{equation}
 \left.I=ie\int\mathcal D\check Q\int\mathcal D\Phi\frac{\delta }{\delta \chi(t)} e^{iS_w+iS_I+iS_\Phi}\right|_{\chi(t)=0}
\label{curgen}
 \end{equation}
one derives the equilibrium supercurrent $I(\theta)$ across our $SNS$ junction.

 \section{Josephson current in the presence of interactions}
Our assumption about the presence of tunnel barriers at $NS$-interfaces with resistances $R_{l,r}$ strongly exceeding $R_N$ allows us to evaluate the
current (\ref{curgen}) perturbatively in tunneling. In the leading order in the tunneling
term $S_I$ (\ref{si}) from Eq. (\ref{curgen}) we obtain
\begin{equation}
  I(t)=\frac{i}{16}\frac{\pi^2}{e^3R_I^lR_I^ra^4}\left\langle \frac{\delta \Tr_r[\check Q_{SC}^{r}\check Q]}{\delta\chi(t)}\Tr_l[\check Q_{SC}^{l}\check Q] \right\rangle_{\check Q,\Phi},
  \label{curr}
\end{equation}
where the averaging is now performed with the action $S_w+S_\Phi$.

In order to proceed we will employ the strategy already developed in Ref. \onlinecite{SZK}.
The averages over the field $\check Q$ will be handled within the Gaussian approximation. To this end we expand the $\check Q$ matrix around the saddle point (\ref{nsp}) as
\begin{equation}
\check Q\approx \check Q_N+i\check Q_N\circ\check {\mathcal U}\circ\check W\circ\check {\mathcal U}-\frac12 \check Q_N\circ\check {\mathcal U}\circ\check W\circ\check W\circ\check {\mathcal U}+...
\label{parametriz}
\end{equation}
Here the matrix $\check W$ describes the soft modes of the system, diffusons $\hat d$ and Cooperons $\hat c_{1,2}$.  It has the form
\begin{equation}
\check W=\left(\begin{array}{cc}
\hat c_1 & \hat d \\
\hat d^\dag & \hat c_2
\end{array}
\right),\  \hat c_{1,2}=\left(\begin{array}{cc}
0 & c_{1,2} \\
c^\dag_{1,2} & 0
\end{array}\right),
\
\hat d=\left(\begin{array}{cc}
d_1 & 0 \\
0 & d_2
\end{array}\right),
\end{equation}
where $(...)^\dag$ denotes the Hermitean conjugation procedure, i.e. $a^\dag(t,t')\equiv \bar a(t',t)$. Expanding the action $S_w$ (\ref{sw}) up to the second order in these fields one recovers four different contributions
\begin{equation}
S_w=S_w^{(0,2)}+S_w^{(1,2)}+S_w^{(2,1)}+S_w^{(2,2)},
\end{equation}
where the term $S^{(i,j)}$ is proportional to the $i$-th power of the electromagnetic potentials and to the $j$-th power of the matrix $\check W$. By direct calculation one can verify that the term $S^{(2,1)}$ depends only on the diffuson fields
which -- as we will see later -- turn out to be irrelevant for the problem under consideration. Hence, our action does not contain the first power of the Cooperon fields, and the corresponding
propagator -- the Cooperon $\mathcal C$ -- can be obtained as a solution of a linear inhomogeneous equation containing the first and the second powers of the
electromagnetic potentials.

Let us now evaluate the combination $\Tr_{l,r}[\check Q_{SC}^{l}\check Q]$. For this purpose it suffices to retain only the first order terms in $\check W$. After some algebra  with the aid of the parametrization  (\ref{parametriz}) we get
\begin{widetext}
\begin{multline}
-i\Tr[\check Q_{SC}^l\check Q]\approx\Tr[e^{-i\hat{\mathcal K}}\circ\hat g\circ e^{i\hat{\mathcal K}}\circ\hat{\mathcal U}\circ\hat \sigma_z(\hat d_1-\hat\sigma_x\hat d_2^T\hat\sigma_x)\circ \hat{\mathcal U} ]
\\
-\Tr[(\hat a\circ f_R\circ\hat a+\hat b\circ f_R\circ\hat b+F\circ(\hat a\circ f_R\circ\hat b+\hat b\circ f_R\circ\hat a))
\circ\hat\tau_x(\hat c_1-\hat\tau_x\hat c_2^T\hat\tau_x)]
\\-\Tr[(\hat a\circ F-F\circ\hat a+F\circ\hat b\circ F-\hat b)\circ( f_R-f_A)\circ\hat b\circ\hat\tau_x(\hat c_1-\hat\tau_x\hat c_2^T\hat\tau_x)]
\label{trqq}
\end{multline}
where
\begin{equation}
\hat{\mathcal U}(t,t')=\left(\begin{array}{cc}\delta(t-t'-0) &
F(t,t') \\ 0 &
-\delta(t-t'+0)\end{array}\right),\quad \hat g =\hat{\mathcal U}\circ\left(\begin{array}{cc}
g_R &0 \\
0 & g_A
\end{array}
\right)\circ\hat{\mathcal U}, \quad \hat{\mathcal K}=\left(\begin{array}{cc}
\mathcal K^+ &\mathcal K^-\\
\mathcal K^- &\mathcal K^+
\end{array}
\right),\quad \hat d_{1,2}=\left(\begin{array}{cc}
0 & d_{1,2} \\
d^\dag_{1,2} & 0
\end{array}\right)
\end{equation}
\end{widetext}
and $\hat a=e^{-i\mathcal K^+\hat\sigma_z}\cos(\mathcal K^-)$, $\hat b=-i\hat\sigma_ze^{-i\mathcal K^+\hat\sigma_z}\sin(\mathcal K^-)$. Eq. (\ref{trqq}) applies in the first order in both the Cooperon and the diffuson fields and contains three different contributions. The one in the first line of Eq. (\ref{trqq}) is proportional to the diffusion field being independent of the superconducting phase $\theta$. For this reason such a term is irrelevant for the Josephson current and it will be disregarded below. The contribution in the last line of Eq. (\ref{trqq}) is proportional to the difference between the retarded and advanced anomalous Green functions. In the non-interacting limit this contribution vanishes identically while in the presence of electron-electron interactions it differs from zero only at energies above the superconducting gap. In other words, this contribution is caused by quasiparticles excited by the fluctuating electromagnetic fields mediating such interactions. Clearly, such quasiparticle contribution can be neglected in the low temperature limit considered here.
Hence, we can restrict our analysis only to terms in the second line of Eq. (\ref{trqq}). Deep in the subgap regime one has
\begin{multline}
\Tr[\check Q_{SC}^l\check Q]\approx-i\sqrt{2}\Tr[(F\circ(\hat a\circ f_R\circ\hat b+\hat b\circ f_R\circ\hat a)\\+
\hat a\circ f_R\circ\hat a+\hat b\circ f_R\circ\hat b)
\circ\hat\tau_x\hat c_{as}],
\label{TrQ}
\end{multline}
where we defined
\begin{equation}
\hat c_{as}=(\hat c_1-\hat\tau_x\hat c_2^T\hat\tau_x)/\sqrt{2}.
\label{cas}
\end{equation}
Note that in the non-interacting limit the above expressions eventually reduce to the well-known result \cite{ALO}, see Appendix A for the corresponding analysis.

It is instructive to look at the spin structure of the combination (\ref{cas}). Since the field $\hat c_1$ ($\hat c_2$) corresponds to $\uparrow\downarrow$ ($\downarrow\uparrow$) configuration (as it is also illustrated in Fig. 1), it is easy to observe that $\hat c_{as}$ (\ref{cas}) accounts for the antisymmetric singlet combination $(\uparrow\downarrow-\downarrow\uparrow)/\sqrt{2}$, which is nothing but the spin structure of a Cooper pair in a conventional superconductor. In this respect the Cooperon fields relevant here are markedly different from those encountered, e.g., within the weak localization problem \cite{GZ1,GZ3,GZ4,GZS,GZ5} described either by $\uparrow\uparrow$ or by $\downarrow\downarrow$ spin configurations, see \cite{SZK} for more details on this issue.

One can distinguish two different contributions describing interaction effects. The first one contains the $\mathcal K$ field at either one of the two $NS$ interfaces. This contribution is encoded in the matrices $\hat a$, $\hat b$ and accounts for uniform in space fluctuations of the electromagnetic field in the $N$-metal representing, e.g., Coulomb blockade effects. Such fluctuations can be handled exactly, see Appendix B for further details.

The second contribution is controlled by the $\Phi_{\mathcal K}$ and ${\bm A}_{\mathcal K}$ fields and includes non-uniform in space electromagnetic fluctuations in the bulk of the normal metal. This contribution can be expressed via the propagator of the Cooperon field and, as we will demonstrate below, it is responsible for dephasing of Cooper pairs inside the $N$-metal.

\section{Dephasing of the Josephson current}

At sufficiently low temperature and in the absence of interactions Cooper pairs entering the normal metal from a superconductor can diffuse at a very long distance
without losing their coherence. However, in the presence of interactions the wave function of a propagating Cooper pair accumulates an extra randomly fluctuating phase which eventually yields destruction of quantum coherence at length scales exceeding the so-called decoherence length $L_\varphi$ which remains finite down to zero temperature \cite{SZK}.

In order to analyze this effect one can employ different approximations. In the limit
of sufficiently short $SNS$ junctions $L \ll L_\varphi$ one can proceed perturbatively
in the interactions which amounts to formally expanding the exponent in Eq. (\ref{curgen}) in powers of the $\Phi_{\mathcal K},{\bm A}_{\mathcal K}$ fields and to making use of the Wick's theorem. In the case of $NS$ structures it was demonstrated \cite{SZK,SZ14} that this expansion yields non-zero dephasing rate of Cooper pairs at $T=0$ already in the first order.

On the other hand, in the most interesting case
of longer $SNS$ junctions with $L \gtrsim L_\varphi$ (which we merely address here)
this approach is clearly insufficient. An appropriate approximation in the latter case
is the semiclassical expansion of the effective action in the fluctuating fields $\Phi_{\mathcal K}^-,{\bm A}_{\mathcal K}^-$ which allows to correctly analyze the effect of quantum dephasing of Cooper pairs. This approach will be employed below in this section.

In the lowest (zero) order in the "quantum" fields $\Phi_{\mathcal K}^-,{\bm A}_{\mathcal K}^-$ for the combination (\ref{TrQ}) one finds
\begin{widetext}
\begin{multline}
\Tr[\check Q_{SC}^l\check Q]\approx
-i\sqrt{2}\int_l d^{d-1}{\bm r}_l\int dtdt'\int\frac{d\varepsilon d\omega_1d\omega_2}{(2\pi)^3} e^{i\omega_1t+i\omega_2t'}\frac{\Delta}{\sqrt{(\varepsilon+i0)^2-\Delta^2}}
\\\times(e^{i(\mathcal K^+({\bm r}_l,t)+\mathcal K^+({\bm r}_l,t'))} c_{as}({\bm r}_l,\varepsilon-\omega_2,\varepsilon+\omega_1)+e^{-i(\mathcal K^+({\bm r}_l,t)+\mathcal K^+({\bm r}_l,t'))}\bar c_{as}({\bm r}_l,\varepsilon+\omega_1,\varepsilon-\omega_2)),
\end{multline}
\begin{multline}
\frac{\delta\Tr[\check Q_{SC}^r\check Q]}{\delta\chi(t)}\approx
\frac{1}{\sqrt{2}}\int_r d^{d-1}{\bm r}_r\int dt'\int\frac{d\varepsilon d\omega_1d\omega_2}{(2\pi)^3}\left( e^{i\omega_1t+i\omega_2t'}+e^{i\omega_2t+i\omega_1t'}\right)\frac{\Delta}{\sqrt{(\varepsilon+i0)^2-\Delta^2}}\tanh\left(\frac{\varepsilon+\omega_1}{2T}\right)
\\\times (e^{-i(\mathcal K^+({\bm r}_r,t)+\mathcal K^+({\bm r}_r,t')-i\theta)} \bar c_{as}({\bm r}_r,\varepsilon+\omega_1,\varepsilon-\omega_2)-e^{i(\mathcal K^+({\bm r}_r,t)+\mathcal K^+({\bf r}_r,t')-i\theta)}c_{as}({\bf r}_r,\varepsilon-\omega_2,\varepsilon+\omega_1) ).
\end{multline}
In order to evaluate the current across our system it is necessary to perform averaging over the Cooperon fields $c_{as}$ in Eq. (\ref{curr}) as well as to integrate
over the electromagnetic fields $\Phi_{\mathcal K}^+$ and ${\bm A}_{\mathcal K}^+$. Let us introduce the Cooperon propagator $\mathcal C_{as}$ defined by means of the equation
\begin{equation}
 \langle \bar c_{as}({\bm r}_1,\varepsilon_1,\varepsilon_1') c_{as}({\bm
r}_2,\varepsilon_2,\varepsilon_2')
\rangle=\frac{2}{\pi\nu}\int dtd\tau d\tau' e^{-i(\varepsilon_1-\varepsilon_1'-\varepsilon_2+\varepsilon_2')t+i(\varepsilon_1+\varepsilon_1')\tau/2-i(\varepsilon_2+\varepsilon_2')\tau'/2}    \mathcal C_{as}({\bm r}_1,{\bm r}_2;\tau,\tau';t).
\end{equation}
This propagator is a functional of the electromagnetic potentials. It satisfies the following diffusion-like  equation
\begin{multline}
\left( 2\partial_\tau-i\Phi_{\mathcal K}^+({\bm r},t-\tau/2)+i\Phi_{\mathcal K}^+({\bm r},t+\tau/2)-D(\partial_{\bm r}+i{\bm A}^+_{\mathcal K}({\bm r},t-\tau/2)+i{\bm A}^+_{\mathcal K}({\bm r},t+\tau/2))^2 \right)\mathcal C_{as}({\bm r},{\bm r}';\tau,\tau';t)\\=\delta({\bm r}-{\bm r}')\delta(\tau-\tau').
\label{coop2}
\end{multline}

Combining all the above equations one can express the Josephson current $I$ in terms of the Cooperon propagator $\mathcal C_{as}$. We obtain
\begin{equation}
I=\frac{\pi T\Delta^2\sin\theta}{e^3\nu R_I^rR_I^la^4}\int_l d^{d-1}{\bm r}_l
\int_r d^{d-1}{\bm r}_r\int\limits_0^\infty d\tau\int\limits_0^\infty d\tau'\int\limits_{-\infty}^\infty dt
\frac {J_0(\Delta\tau)J_0(\Delta\tau')}{\sinh(2\pi T t)}\mathcal P({\bm r}_r,{\bm r}_l;t;\tau,\tau'),
\label{currgen}
\end{equation}
where
\begin{equation}
\mathcal P({\bm r}_r,{\bm r}_l;t;\tau,\tau')=\left\langle e^{-i(\mathcal K^+({\bm r}_r,t+\tau/2)+\mathcal K^+({\bm r}_r,t-\tau/2)-\mathcal K^+({\bm r}_l,\tau'/2)-\mathcal K^+({\bm r}_l,-\tau'/2))} \mathcal C_{as}({\bm r}_r,{\bm r}_l;2t-\tau,\tau';0)\right\rangle_{\Phi}.
\end{equation}
Here the effect of electron-electron interactions is encoded in the function $\mathcal P({\bm r}_r,{\bm r}_l;T;\tau,\tau')$ which can be rewritten as a path integral over diffusive trajectories
\begin{multline}
\mathcal P({\bm r}_r,{\bm r}_l;t;\tau,\tau')=\frac12\Theta(t-(\tau+\tau')/2)\\\times\left\langle e^{i(\mathcal K^+({\bm r}_r,-t+\tau/2)-\mathcal K^+({\bm r}_r,t+\tau/2))}
\int\limits^{{\bm x}(2t-\tau)={\bm r}_r}_{{\bm x}(\tau')={\bm r}_l}\mathcal D{\bf x}(t')e^{-\int\limits_{\tau'}^{2t-\tau} dt'
\left(\frac{(\dot{\bm x}(t'))^2}{2D}-\frac{i}{2}(\Phi^+({\bm x}(t'),-t'/2)-\Phi^+({\bm x}(t'),t'/2)\right)}
\right\rangle_\Phi,
\end{multline}
where $\Theta (t)$ is the Heavyside step function. This representation is convenient for averaging over the field $\Phi$. As a result we obtain
\begin{equation}
\mathcal P({\bm r}_r,{\bm r}_l;t;\tau,\tau')=\frac12\Theta(t-(\tau+\tau')/2)\int\limits^{{\bm x}(2t-\tau)={\bm r}_r}_{{\bm x}(\tau')={\bm r}_l}\mathcal D{\bm x}(t')e^{-\int\limits_{\tau'}^{2t-\tau} dt'
\frac{(\dot{\bm x}(t'))^2}{2D}-S_{{ int}}[{\bm x}(t')]},
\label{pgen}
\end{equation}
where
\begin{multline}
S_{int}[{\bm x}(t)]=\frac{i}{2}\left(\mathcal V^{++}_{\mathcal K}({\bm r}_r,{\bm r}_r,0)-\mathcal V^{++}_{\mathcal K}({\bm r}_r,{\bm r}_r,2t)\right)\\
+\frac{i}{4}\int\limits_{\tau'}^{2t-\tau} dt' \left( \mathcal V^{++}_{\mathcal K\Phi}({\bm r}_r,{\bm x}(t'),-t+(\tau+t')/2)-\mathcal V^{++}_{\mathcal K\Phi}({\bm r}_r,{\bm x}(t'),-t+(\tau-t')/2)
\right.\\\left.-\mathcal V^{++}_{\mathcal K\Phi}({\bm r}_r,{\bm x}(t'),t+(\tau+t')/2)+\mathcal V^{++}_{\mathcal K\Phi}({\bm r}_r,{\bm x}(t'),t+(\tau-t')/2) \right)
\\+\frac{i}{4}\int\limits_{\tau'}^{2t-\tau} dt'\int\limits_{\tau'}^{t'} dt''\left(\mathcal V^{++}_{\Phi}({\bm x}(t'),{\bm x}(t''),(t'-t'')/2)-\mathcal V^{++}_{ \Phi}({\bm x}(t'),{\bm x}(t''),(t'+t'')/2)\right).
\label{sintgen}
\end{multline}
\end{widetext}
Here $\mathcal V^{++}_{\Phi}({\bm r},{\bm r}',t-t')=-2i\langle\Phi^+({\bm r},t)\Phi^+({\bm r}',t')\rangle_\Phi$ and the functions $\mathcal V^{++}_{\mathcal K\Phi}$, $\mathcal V^{++}_{\mathcal K}$ are defined analogously. Eq. (\ref{currgen}) combined with Eqs. (\ref{pgen}) and (\ref{sintgen}) constitutes the general expression for the Josephson current suitable for further analysis of quantum dephasing by electron-electron interactions.

A standard (and sufficient for our purposes) approximation in Eq. (\ref{pgen}) amounts to replacing
$\langle\langle e^{-S_{int}}\rangle\rangle\approx e^{-\langle\langle S_{int}\rangle\rangle}$, where $\langle\langle...\rangle\rangle$ implies
averaging over diffusive electron trajectories. It is convenient to introduce the dephasing function $\mathcal I$ as
\begin{equation}
\mathcal P({\bm r}_r,{\bf r}_l;t;0,0)=\mathcal D({\bm r}_r,{\bm r}_l;t)e^{-\mathcal I({\bm r}_r,{\bm r}_l;t)},
\end{equation}
where
\begin{widetext}
\begin{multline}
\mathcal I({\bm r}_r,{\bm r}_l;t)=\frac{i}{2}\left(\mathcal V^{++}_{\mathcal K}({\bm r}_r,{\bm r}_r,0)-\mathcal V^{++}_{\mathcal K}({\bm r}_r,{\bm r}_r,2t)\right)
+\frac{i}{2}\int d^d{\bm x}\int\limits_{0}^{2t} dt' \left( \mathcal V^{++}_{\mathcal K\Phi}({\bm r}_r,{\bm x},-t+t'/2)-\mathcal V^{++}_{\mathcal K\Phi}({\bm r}_r,{\bm x},-t-t'/2)\right.\\\left.
-\mathcal V^{++}_{\mathcal K\Phi}({\bm r}_r,{\bm x},t+t'/2)+\mathcal V^{++}_{\mathcal K\Phi}({\bm r}_r,{\bm x},t-t'/2) \right)\frac{\mathcal D({\bm r}_r,{\bm x};2t-t')\mathcal D({\bm x},{\bm r}_l;t')}{\mathcal D({\bm r}_r,{\bm r}_l;2t)}
\\+i\int d^d{\bm x}d^d{\bm x}'\int\limits_{0}^{2t} dt'\int\limits_{0}^{t'} dt''\left(\mathcal V^{++}_{\Phi}({\bm x},{\bm x}',(t'-t'')/2)-\mathcal V^{++}_{ \Phi}({\bm x},{\bm x}',(t'+t'')/2)\right)\\\times
\frac{\mathcal D({\bm r}_r,{\bm x};2t-t')\mathcal D({\bm x},{\bm x'};t'-t'')\mathcal D({\bm x}',{\bm r}_l;t'')}{\mathcal D({\bm r}_r,{\bm r}_l;2t)}.
\label{deff}
\end{multline}
\end{widetext}
As we already pointed out, in the long junction limit $\varepsilon_{Th}=\pi^2D/L^2\ll \Delta$ considered here
the expression for the Josephson current is dominated by the times exceeding the inverse Thouless energy $1/\varepsilon_{Th}$.
Accordingly, it suffices to establish only the leading time behavior of this expression, which can be derived from the analysis of the most singular terms
of its Fourier transform.

The correlators of the electromagnetic potentials in the normal metal have the form
\begin{equation}
\mathcal V^{++}_\Phi({\bm r},{\bm r'},\omega)\approx -i\omega \coth\left(\frac{\omega}{2T}\right)\sum\limits_{n=1}^{E_n<1/l^2} \frac{\psi_n({\bm r}) \psi_n({\bm r}')}{\nu DE_n},
\end{equation}
\begin{multline}
\mathcal V^{++}_{\mathcal K}({\bm r},{\bm r'},\omega)\approx -i\omega \coth\left(\frac{\omega}{2T}\right)\\\times\sum\limits_{n=1}^{E_n<1/l^2} \frac{\psi_n({\bm r}) \psi_n({\bm r}')}{\nu DE_n ((DE_n)^2+\omega^2)},
\label{Vk}
\end{multline}
and $\mathcal V^{++}_{\mathcal K\Phi}({\bm r},{\bm r'},\omega)\approx 0$ in the so called ``universal limit'' of strong interactions.
Here $E_n=\pi^2 n^2/L^2$ and  $\psi_n({\bm x})=\sqrt\frac{2}{(1+\delta_{n,0})L\Gamma}\cos\left(\frac{\pi n x}{L}\right)$ are the eigenvalues and the
eigenfunctions of the Laplace operator with the von Neumann boundary conditions. Our analysis of the dephasing function reveals that at sufficienly long times
it is dominated by the last integral in Eq. (\ref{deff}), whereas the first term in that equation just provides a constant which
cannot be determined by means this approach. This constant, however, can be conveniently recovered by treating the short wire limit in which case
the effect of interactions on the Cooperon can be neglected and one can set $\Phi_{\mathcal K}$ equal to zero. The algebra remains the same and
now amounts to substituting
\begin{widetext}
\begin{equation}
\mathcal P({\bm r}_r,{\bm r}_l;t;\tau,\tau')\to \mathcal P_0({\bm r}_r,{\bm r}_l;t;\tau,\tau')=\left\langle e^{-i(\mathcal K^+({\bm r}_r,t+\tau/2)+\mathcal K^+({\bm r}_r,t-\tau/2)-\mathcal K^+({\bm r}_l,\tau'/2)-\mathcal K^+({\bm r}_l,-\tau'/2))}\right\rangle_{\Phi}
\mathcal D({\bm r}_r,{\bm r}_l;t-(\tau+\tau')/2).
\end{equation}
Evaluating the average in a standard manner, we obtain
\begin{equation}
\mathcal I(0,L;t)\approx -\log(\mathcal A)
 -\frac{t}{2\nu La^2}\int\frac{dz}{2\pi}\coth\left(\frac{z}{2T}\right)\frac{1-(1-i)L\sqrt{\frac{z}{2D}}
 \coth\left((1-i)L\sqrt{\frac{z}{2D}}\right))}{z},
\label{defres}
\end{equation}
\end{widetext}
where the integral is interpreted as a principal value at small $z$ and, as usually, it should be cut off at the largest energy scale
of the inverse elastic time $\sim \tau_e^{-1}$.
The time-independent constant
\begin{equation}
\mathcal A\approx\begin{cases}
(\varepsilon_{Th}\tau_{RC})^{\frac{8}{3g}} e^{-\frac{6.105}{g}-\frac{8\pi T}{3g\varepsilon_{Th}}}, & T\ll \varepsilon_{Th},\\

(T\tau_{RC})^{\frac{8}{3g}} e^{-\frac{2.892}{g}-\frac{8\pi T}{3g\varepsilon_{Th}}}, &  \varepsilon_{Th}\ll T
\end{cases}
\label{mA}
\end{equation}
depends on the dimensionless conductance of the normal wire $g=4\pi\nu D a^2/L$ as well as on the corresponding
$RC$-time $\tau_{RC}=\frac{L^2C}{2\nu e^2Da^2}$ (where $C$ denotes the capacitance per unit wire length),
which we will assume to be short further below. Evaluating the integral in Eq. (\ref{defres}), in the low temperature limit $T\ll\varepsilon_{Th}$ one finds
\begin{equation}
\mathcal I(0,L;t)\approx -\log(\mathcal A)
 +\frac{t}{\tau_\varphi},
\end{equation}
where the inverse dephasing time equals to
\begin{equation}
\frac{1}{\tau_\varphi}\simeq \frac{1}{4\pi\nu a^2\sqrt{2D\tau_e}}-\frac{\log\left(\frac{2L^2}{D\tau_e}\right)}{4\pi\nu a^2 L}+\frac{\pi L^3 T^2}{270\nu D^2 a^2}+...
\label{tauphi}
\end{equation}

With the aid of all the above expressions it is now straightforward to derive the Josephson current taking into account the effect of Cooper pair
dephasing by electron-electron interactions. We obtain
\begin{equation}
I=\frac{\pi T\mathcal A\sin\theta}{2e^3\nu R_I^rR_I^lLa^2}
\int\limits_{0}^\infty dt
\frac {\vartheta_3\left(\frac{1}{2},e^{-\frac{\pi^2 D t}{L^2}}\right)e^{-\frac{t}{\tau_\varphi}}}{\sinh(2\pi T t)},
\end{equation}
where $\vartheta_k(z,q)$ is the $k$-th Jacobi theta function. One observes that the Josephson current -- as compared to the non-interacting limit (\ref{currni}) --
essentially depends on the extra energy scale which is the inverse dephasing time $1/\tau_\varphi$.

Provided the temperature is sufficiently high $L_T=\sqrt{D/2\pi T}\lesssim L$
the Josephson current reduces to an exponentially small value
\begin{equation}
I=\frac{2 \pi T\mathcal AL_c\sin\theta }{e^3\nu D R_I^rR_I^l a^2 }e^{-\frac{L}{L_c}}.
\end{equation}
where
\begin{equation}
L_c=\sqrt{\frac{D\tau_\varphi}{1+2\pi T\tau_\varphi}}
\label{Lc}
\end{equation}
 defines the critical length which -- unlike in the non-interacting case -- now depends on both temperature and the dephasing time $\tau_\varphi$.

In the opposite low temperature limit $L,L_\varphi \ll L_T$ one finds
\begin{equation}
I\simeq\frac{\mathcal A\sin\theta}{2e^3\nu R_I^rR_I^lLa^2}\log\left(\coth\left(\frac{L}{2\sqrt{D\tau_\varphi}}\right)\right).
\label{Ilog}
\end{equation}

\section{Discussion}

The above results clearly demonstrate that dephasing of Cooper pairs by electron-electron interactions may
strongly influence the Josephson current in diffusive $SNS$ junctions at low temperatures. The supercurrent suppression
in such structures is controlled by the ratio of the normal wire length $L$ to the effective critical length
$L_c$ (\ref{Lc}). Note that the latter parameter can also be rewritten as
\begin{equation}
 L_c=\frac{L_TL_\varphi}{\sqrt{L_T^2+L_\varphi^2}},
\end{equation}
where we defined the Cooper pair dephasing length $L_\varphi =\sqrt{D\tau_\varphi}$. In the low temperature limit
$L_T \gg L_\varphi$ the magnitude of the Josephson current depends on the relation between the two lengths $L$ and
$L_\varphi$. In this limit and for $L \ll L_\varphi$ this current is not significantly affected by electron-electron interactions,
i.e. $I$ drops almost linearly with $1/L$ and depends on $L_\varphi$ only logarithmically, cf. Eq. (\ref{Ilog}).
In this case non-vanishing Cooper pair dephasing provides a natural cutoff of the divergence
in Eq. (\ref{alo}) at $T \to 0$ \cite{FN}. On the other hand, as soon as the length $L$ exceeds $L_\varphi$ the power law
dependence of $I$ on $L$ turns into an exponential one $I \propto \exp (-L/L_\varphi )$. Thus, in sufficiently
long $SNS$ junctions the Josephson current is exponentially suppressed even at $T=0$ due to the effect of
dephasing of Cooper pairs which occurs in the $N$-metal in the presence of electron-electron interactions.

The length $L_\varphi$ constitutes a new fundamental parameter in our problem which can be detected experimentally
\cite{FN2}
by measuring the low temperature Josephson critical current in diffusive $SNS$ junctions as a function of the normal wire length $L$,
see also Fig. 2. In fact, such kind of experiments were recently performed \cite{MM} and their results appear to
be consistent with our theoretical predictions. A complementary way to experimentally probe Cooper pair dephasing
in sufficiently long $SNS$ junctions is to measure the temperature dependence of the supercurrent $I(T)$ which
should crossover between the interaction-dominated regime
$I \propto \exp (-L/L_\varphi )$ at $T\tau_\varphi \lesssim 1$ and the high temperature one  $T\tau_\varphi \gg 1$
in which case electron-electron interactions are irrelevant and the standard dependence $I \propto \exp (-L/L_T )$
is realized, see also Fig. 3.

\begin{figure}[t]
\includegraphics[width=0.95\columnwidth]{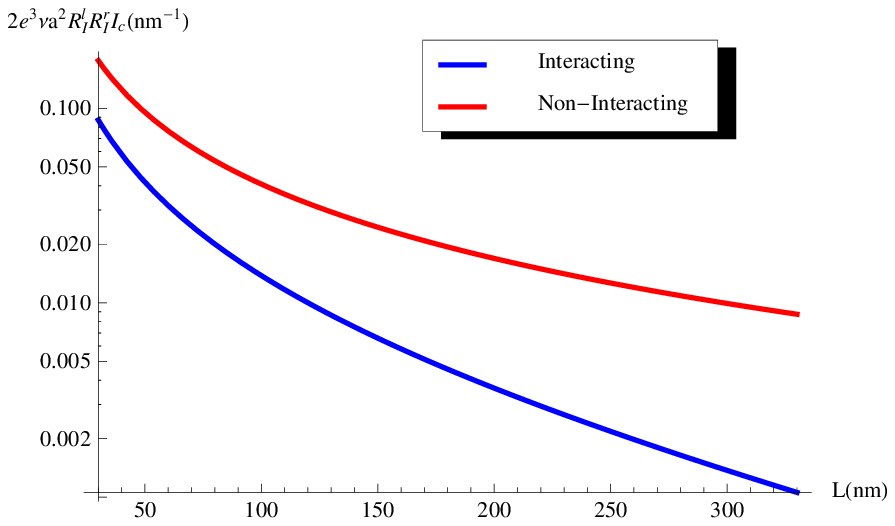}
\caption{Josephson current in diffusive $SNS$ junctions as a function of the normal metal wire length $L$ at $T=1{\rm mK}$. Here we set
$D=20 {\rm cm^2/s}$, $a=10{\rm nm}$ and $L_\varphi=215{\rm nm}.$}
\end{figure}
\begin{figure}[t]
\includegraphics[width=0.95\columnwidth]{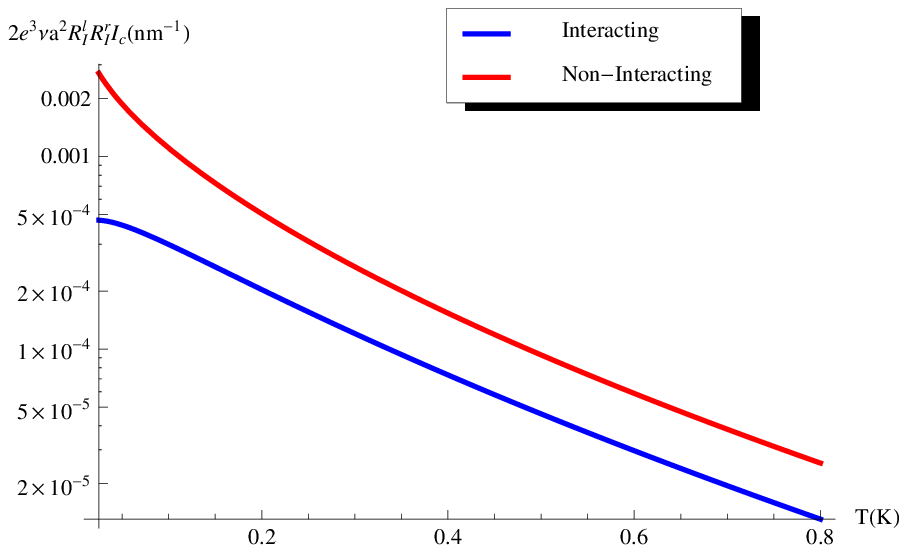}
\caption{Josephson current in diffusive $SNS$ junctions as a function of temperature for $L_T =\sqrt{D/(2 \pi T)}\lesssim L=400 {\rm nm}$ with and without interaction. The parameters are the same as in Fig. 2.}
\end{figure}

Let us also note that an additional interaction-induced suppression of the Josephson current is encoded in the parameter ${\mathcal A}$ (\ref{mA}). This is a specific contribution to dephasing of Cooper pairs provided by uniform in space fluctuations of the electromagnetic field \cite{SZK,Artem}. The magnitude of this effect is controlled
by the dimensionless conductance of the normal wire $g$. As this parameter typically remains large for generic metallic junctions, the corresponding reduction of the supercurrent may be less significant than that caused by non-uniform in space electromagnetic fluctuations giving rise to the parameter $\tau_\varphi$ (\ref{tauphi}).

It is necessary to emphasize that both the dephasing time $\tau_\varphi$ and the
dephasing length $L_\varphi$ derived here in the limit $T \to 0$ coincide -- up to a numerical factor of order one -- with analogous parameters previously obtained from the calculations of the subgap (Andreev) conductance of $NS$-structures \cite{SZK} and
of the weak localization correction to the conductance of normal metals \cite{GZ1,GZ3,GZ4,GZS,GZ5}. This agreement is, of course, by no means a pure coincidence. Rather it emphasizes universality of the phenomenon of low temperature quantum decoherence by electron-electron interactions which can be observed in a variety of normal and hybrid normal-superconducting structures. The underlying physics of the effect is simple and remains essentially the same in all situations.  Here two electrons initially forming a Cooper pair propagate in the normal metal between two
superconductors, pick up random phases while interacting with the fluctuating electromagnetic field produced by other electrons and eventually become incoherent
at length scales exceeding $L_\varphi$.

At the same time,
an important peculiar feature of our present problem is that -- unlike in a number of other cases \cite{SZK,GZ1,GZ3,GZ4,GZS,GZ5} -- it addresses {\it non-dissipative} electron transport demonstrating that quantum dephasing of Cooper pairs occurs exactly in the equilibrium ground state of our system. This property of diffusive $SNS$ hybrids is generic, i.e. it is not specific, e.g., to the limit $R_l,R_r\gg R_N$ analyzed here but should also hold for structures with highly transparent inter-metallic interfaces.

\vspace{0.5cm}

\centerline{\bf Acknowledgements}
We acknowledge useful discussions with A.V. Galaktionov. This work was supported in part by RFBR grant 12-02-00520-a.

\appendix

\section{}
Let us briefly demonstrate how to recover the well known results \cite{ALO} for the non-interacting limit by means of our technique. For this purpose it is necessary to simply drop the fluctuating electromagnetic potentials from the above expressions. This step amounts to substituting
\begin{multline}
\mathcal P({\bm r}_r,{\bm r}_l;t;\tau,\tau')\to \mathcal D({\bm r}_r,{\bm r}_l;t-(\tau+\tau')/2)\\=\frac12\Theta(t-(\tau+\tau')/2)\int\limits^{{\bf m}(2t-\tau)={\bm r}_r}_{{\bm x}(\tau')={\bm r}_l}\mathcal D{\bm x}(t')e^{-\int\limits_{\tau'}^{2t-\tau} dt'
\frac{(\dot{\bm x}(t'))^2}{2D}}
\label{dpi}
\end{multline}
in Eq. (\ref{currgen}). Evaluating the path integral in Eq. (\ref{dpi}) for a quasi-one-dimensional normal metallic wire (with length $L$ and cross section $a^2$) and expressing the result via the Jacobi theta function $\vartheta_3(z,q)$, we obtain
\begin{multline}
\mathcal D({\bf x},{\bf y};t)=\frac{\Theta(t)}{4La^2}\left(
\vartheta_3\left(\frac{x+y}{2L},e^{-\frac{\pi^2D t}{2L^2}}\right)\right.\\\left.+\vartheta_3\left(\frac{x-y}{2L},e^{-\frac{\pi^2D t}{2L^2}}\right)\right).
\end{multline}
As a result we arrive at the Josephson current in the form
\begin{multline}
I=\frac{\pi T\Delta^2\sin\theta}{2e^3\nu R_I^rR_I^lLa^2}\int\limits_0^\infty d\tau\int\limits_0^\infty d\tau'
\\\times\int\limits_{\frac{\tau+\tau'}{2}}^\infty dt
\frac {J_0(\Delta\tau)J_0(\Delta\tau')}{\sinh(2\pi T t)}\vartheta_3\left(\frac{1}{2},e^{-\frac{\pi^2 D(2t-\tau-\tau')}{2L^2}}\right).
\label{nires}
\end{multline}
Having in mind that Bessel functions decay at times $\tau,\ \tau'$ exceeding $\Delta^{-1}$ and the function $\vartheta_3$ is nonzero only for
times larger than the inverse Thouless energy $1/\varepsilon_{Th}$, in the limit of sufficiently long junctions $\varepsilon_{Th}\ll \Delta$ and subgap temperatures $T \ll \Delta$ we can safely neglect set $\tau,\ \tau'$ equal to zero everywhere except in the arguments of the Bessel functions. Then we get
\begin{equation}
I=\frac{\pi T\sin\theta}{2e^3\nu R_I^rR_I^lLa^2}
\int\limits_{0}^\infty dt
\frac {\vartheta_3\left(\frac{1}{2},e^{-\frac{\pi^2D t}{L^2}}\right)}{\sinh(2\pi T t)}.
\label{currni}
\end{equation}
Evaluating the integral in Eq. (\ref{currni}) in high and low temperature limits, we obtain
\begin{multline}
I\approx \frac{\sin\theta}{4e^3\nu  R_I^rR_I^l La^2 }\\\times
\begin{cases}
\frac{4L}{L_T}e^{-L/L_T}, & L\gg L_T,\\
\log\left(\frac{4 D }{\pi L^2 T}\right)+\gamma+\frac{7\pi^2 L^4 T^2}{540 D^2}+..., & L\ll L_T,
\end{cases}
\label{alo}
\end{multline}
where $\gamma = 0.577...$ is the Euler constant and $L_T=\sqrt{D/2\pi T}$ is the temperature length. These expressions reproduce the well known result \cite{ALO}. Note that the current (\ref{alo}) formally diverges in the
zero temperature limit $T \to 0$. In the absence of interactions this divergence can be cured only by taking into account higher order tunneling terms.
In the presence of electron-electron interactions this is not necessary, as the low temperature divergence in Eq. (\ref{alo}) is naturally eliminated
by including the effect of Cooper pair dephasing, cf. Eq. (\ref{Ilog}).

Note that an alternative way to regularize the non-interacting result (\ref{alo}) in the limit $T \to 0$ is to take into account Coulomb blockade effects \cite{BFS}. Within the model adopted here this task requires a separate calculation presented below in Appendix B.

\section{}

In order to fully account for charging effects in the case of relatively short normal metal wires
(with length $L$ shorter that $L_\varphi$) within the framework of our formalism it is necessary to retain
the fields $\mathcal K^+$ and $\mathcal K^-$ simultaneously dropping the fields $\Phi_{\mathcal K}$ and
${\bf A}_{\mathcal K}$. The latter approximation implies that averaging over the Cooperon fields should be performed in the non-interacting limit with
\begin{multline}
 \langle \bar c_{as}({\bf r}_1,\varepsilon_1,\varepsilon_1') c_{as}({\bf
r}_2,\varepsilon_2,\varepsilon_2')
\rangle=(2\pi)^2\delta(\varepsilon_1-\varepsilon_2)\delta(\varepsilon_1'-\varepsilon_2') \\\times
\frac{2}{\pi\nu}\sum\limits_{n=0}^\infty\frac{\psi_n({\bf r}_r)\psi_n({\bf r}_l)}{-i(\varepsilon_1+\varepsilon_1')+D E_n}.
\end{multline}
Then the general expression for the supercurrent reads
\begin{widetext}
\begin{multline}
I=\frac{\pi\Delta^2\sin(\theta)}{e^3\nu R_I^rR_I^la^4}
\int\limits_{\Gamma_l} d^{d-1}{\bf r}_l
\int\limits_{\Gamma_r} d^{d-1}{\bf r}_r\int\limits_0^\infty d\tau\int\limits_0^\infty d\tau'\int\limits_{-\infty}^\infty dt\int\limits_{-\infty}^\infty\frac{d\omega}{2\pi}
 J_0(\Delta\tau)J_0(\Delta\tau')
e^{-i\omega(t-(\tau+\tau')/2)}
I_c({\bf r}_r,{\bf r}_l,\tau+\tau';\omega)
\\\times\left({\rm Re}(\mathcal P_{FF}({\bf r}_r,{\bf r}_l;t;\tau,\tau'))+i\coth\left(\frac{\omega}{2T}\right){\rm Im}(\mathcal  P_{FF}({\bf r}_r,{\bf r}_l;t;\tau,\tau'))\right),
\end{multline}
where
\begin{equation}
I_c({\bf r}_r,{\bf r}_l,\tau+\tau';\omega)=T\sum\limits_{m=0}^\infty \sum\limits_{n=0}^\infty
\frac{\psi_n({\bf r}_r)\psi_n({\bf r}_l)e^{-z_m(\tau+\tau')}}{-i\omega+2z_m+D E_n},
\end{equation}
$z_m=\pi T(2m+1)$ and
\begin{multline}
 \mathcal P_{FF}({\bf r}_r,{\bf r}_l;t;\tau,\tau')=
 \left\langle  e^{i(\mathcal K^+({\bf r}_r,t+\tau/2)+\mathcal K^+({\bf r}_r,t-\tau/2)-\mathcal K^+({\bf r}_l,\tau'/2)-\mathcal K^+({\bf r}_l,-\tau'/2))
}\right. \\\times \left.e^{i(\mathcal K^-({\bf r}_r,t+\tau/2)+\mathcal K^-({\bf r}_r,t-\tau/2))-i(\mathcal K^-({\bf r}_l,\tau'/2)+\mathcal K^-({\bf r}_l,-\tau'/2))}  \right\rangle_\Phi .
\label{B4}
\end{multline}
\end{widetext}
All the integrals here should be understood as a principal value.  As before,
let us restrict our analysis to the well pronounced subgap regime and set both $\tau$ and $\tau'$ equal to zero.
Then Eq. (\ref{B4}) reduces to
\begin{equation}
 \mathcal P_{FF}({\bf r}_r,{\bf r}_l;t)=
 \left\langle  e^{2i(\mathcal K^+({\bf r}_r,t)-\mathcal K^+({\bf r}_l,0)+\mathcal K^-({\bf r}_r,t)-\mathcal K^-({\bf r}_l,0))
} \right\rangle_\Phi
\end{equation}
and one readily finds
\begin{multline}
 \mathcal P_{FF}({\bf r}_r,{\bf r}_l;t)=e^{-2i(\mathcal V_{\mathcal K}^{++}({\bf r}_r,{\bf r}_r,0)-\mathcal V_{\mathcal K}^{++}({\bf r}_r,{\bf r}_l,t))}\\
 \times e^{2i(\mathcal V_{\mathcal K}^{+-}({\bf r}_r,{\bf r}_l,t)+\mathcal V_{\mathcal K}^{+-}({\bf r}_r,{\bf r}_l,-t))}.
\end{multline}
The general expressions for the correlators in the above equation have the form \cite{Serota}
\begin{multline}
\mathcal V_{\mathcal K}^{+-}({\bm r},{\bm r'},\omega)=\\-\sum\limits_{n=0}^{E_n<1/l^2}\frac{\psi_n({\bf r})\psi_n({\bf r'})}{(DE_n-i\omega)(2\nu DE_n-iU_0^{-1}\omega)};
\label{vpm}
\end{multline}
\begin{multline}
\mathcal V_{\mathcal K}^{++}({\bm r},{\bm r'},\omega)=-i\omega\coth\left(\frac{\omega}{2T}\right)
\\\times\sum\limits_{n=0}^{E_n<1/l^2}\frac{2(2\nu+U_0^{-1})DE_n \psi_n({\bf r})\psi_n({\bf r'})}{((DE_n)^2+\omega^2)((2\nu DE_n)^2+U_0^{-2}\omega^2)}.
\label{vpp}
\end{multline}
Here $U_0$ denotes the unscreened Coulomb interaction between electrons. In the quasi-1d geometry considered here one has $U_0=e^2 a^2/C$. We also note that the condition $U_0\nu\sim (e p_Fa)^2/(v_FC)\gg 1$ is usually well satisfied in metallic structures.

In order to evaluate the supercurrent across our $SNS$ structure we need to establish the behavior of the correlation functions at times exceeding the inverse Thouless energy $1/\varepsilon_{Th}\gg\tau_{RC}$. In this limit it suffices to ignore all terms in Eqs. (\ref{vpm}) and (\ref{vpp}) except for one with $n=0$ (where one should
 also account for the contribution from the ion jelly in the normal metal).  Then in the long time limit $\varepsilon_{Th}t\gg1$ one finds
\begin{equation}
 \mathcal V_{\mathcal K}^{+-}(x,y,t)\approx -i\Theta(t)t\varepsilon_C, \qquad  \mathcal V_{\mathcal K}^{++}({\bm r},{\bm r'},t)\approx 0,
\end{equation}
where $\varepsilon_C=e^2/(LC)$ is the charging energy of the normal wire. On the other hand, in the short time limit $t\to 0$ one gets
\begin{multline}
\mathcal V_{\mathcal K}^{++}(0,0,0)\approx-\frac{4i\pi T}{3g\varepsilon_{Th}}\\+\frac{4i}{3g}\begin{cases}
\log\left(\frac{\varepsilon_{Th}\tau_{RC}}{\pi^2}\right)+..., & T\ll\varepsilon_{Th},\\
\log(T\tau_{RC})-2.169+..., & \varepsilon_{Th}\ll T.
\end{cases}
\end{multline}
Combining all the above expressions, we obtain
\begin{multline}
\mathcal P_{FF}(0,L;t)
\\\approx\begin{cases}
(\varepsilon_{Th}\tau_{RC})^{\frac{8}{3g}} e^{-\frac{6.105}{g}-\frac{8\pi T}{3g\varepsilon_{Th}}-2i\varepsilon_C |t|}, & T\ll \varepsilon_{Th},\\
(T\tau_{RC})^{\frac{8}{3g}} e^{-\frac{2.892}{g}-\frac{8\pi T}{3g\varepsilon_{Th}}-2i\varepsilon_C |t|}, &  \varepsilon_{Th}\ll T.
\end{cases}
\end{multline}
At high enough temperatures $T\gg \varepsilon_C$ charging effects can be safely neglected and one can set ${\rm Im} \mathcal P_{FF}\sim \sin(2\varepsilon_C |t|)\approx 0$. In the opposite low temperature limit $T\ll\varepsilon_C$ one finds
\begin{multline}
 I=\frac{ \pi T^2\mathcal A\sin\theta}{e^3\nu R_I^rR_I^lLa^2}
 \\\times\sum\limits_{m,k=0}^\infty \sum\limits_{n=-\infty}^\infty \frac{(-1)^n}{z_m+z_k+DE_n}\frac{4\varepsilon_C}{(z_m-z_k)^2+4\varepsilon_C^2}.
\end{multline}
Performing the summation over $n$ we arrive at the result
\begin{multline}
 I=\frac{ \pi T^2\mathcal A\sin\theta}{e^3\nu R_I^rR_I^lDa^2}\sum\limits_{m,k=0}^\infty\sqrt{\frac{D}{z_m+z_k}}\\
 \times\frac{4\varepsilon_C}{((z_m-z_k)^2+4\varepsilon_C^2)
 \sinh\left(L\sqrt{\frac{z_m+z_k}{D}}\right)}.
\label{B13}
\end{multline}
In the limit $T \to 0$ one can replace the double sum in Eq. (\ref{B13}) by the double integral and get
\begin{equation}
I\approx \frac{\mathcal A\sin\theta}{4e^3\nu  R_I^rR_I^l La^2 }\begin{cases}
\log (2\varepsilon_{Th}/(\pi^2\varepsilon_{C})), &  \varepsilon_{Th}\gg\varepsilon_C,\\
0.271\varepsilon_{Th}/\varepsilon_{C}, &\varepsilon_{Th}\ll\varepsilon_C.
\end{cases}
\label{aloreg}
\end{equation}
In the limit $\varepsilon_{Th}\gg\varepsilon_C$ the above expression holds within the logarithmic accuracy and demonstrates that Coulomb blockade effects
naturally eliminate the divergence of the non-interacting result (\ref{alo}). A similar observation
was previously made \cite{BFS} within a simple model taking into account both the gate capacitance and
those of the tunnel barriers. Although we deliberately ignored all these capacitances here, if needed, they
can easily be restored by a proper modification of the expressions for the correlators  (\ref{vpm}), (\ref{vpp}).


\begin{references}
\bibitem{FN} In order to correctly evaluate the cutoff time under the logarithm in Eq.
(\ref{Ilog}) in the short junction limit $L \ll L_\varphi$ it can be more appropriate to
use the perturbation theory in the interaction \cite{SZK,SZ14}.
\bibitem{FN2} Strictly speaking, it is necessary to distinguish Cooper pair dephasing by electron-electron interactions
from an alternative dephasing mechanism of electron scattering on magnetic impurities which could also be present in the normal wire.
One can demonstrate \cite{GR} that the corresponding dephasing rate enters into the expression for $L_c$ (\ref{Lc})
in exactly the same way as $1/\tau_\varphi$ derived here. On the other hand, the dephasing time due
to magnetic impurities \cite{GR} is entirely different from that caused by electron-electron interactions, thus
enabling one to reliably distinguish these two dephasing mechanisms in modern experiments.
\end{references}
\end{document}